\documentclass{elsart3}

\usepackage{graphicx}

\usepackage{amssymb}

\begin{document}

\begin{frontmatter}



\title{Mixed state microwave resistivity of cuprate 
superconductors}


\author{N. Pompeo\thanksref{uni3}},
\ead{pompeo@fis.uniroma3.it}
\author{L. Muzzi\thanksref{uni3}\thanksref{PresentAddress}},
\author{S. Sarti\thanksref{uni1}},
\author{R. Marcon\thanksref{uni3}},
\author{R. Fastampa\thanksref{uni1}},
\author{M. Giura\thanksref{uni1}},
\author{M. Boffa\thanksref{salerno}},
\author{M.C. Cucolo\thanksref{salerno}},
\author{A.M. Cucolo\thanksref{salerno}},
\author{C. Camerlingo\thanksref{cnr}}, and
\author{E. Silva\thanksref{uni3}}.
\address[uni3]{Dipartimento di Fisica ''E. Amaldi'' and INFM, Universit\`a di Roma Tre, Via della Vasca Navale 84, I-00146 Roma, Italy.}
\address[uni1]{Dipartimento di Fisica and INFM, Universit\`a di Roma ''La Sapienza'', I-00185 Roma, Italy.}
\address[salerno]{Dipartimento di Fisica and INFM, Universit\`a di Salerno, Baronissi, Salerno, Italy.}
\address[cnr]{CNR - Istituto di Cibernetica ''E. Caianiello'', Napoli, Italy.}
\thanks[PresentAddress]{Present address: ENEA, Frascati, Italy.}
\begin{abstract}
We present a compared experimental investigation of the $(a,b)$ plane
vortex-state complex resistivity at 48 GHz in
YBa$_{2}$Cu$_{3}$O$_{7-\delta}$, SmBa$_{2}$Cu$_{3}$O$_{7-\delta}$ and
Bi$_{2}$Sr$_{2}$CaCu$_{2}$O$_{8+x}$.  In
YBa$_{2}$Cu$_{3}$O$_{7-\delta}$ and SmBa$_{2}$Cu$_{3}$O$_{7-\delta}$
the field dependence of the response can be consistently described by
a combination of flux flow and strong pair breaking due to the
presence of lines of nodes in the gap.  In
Bi$_{2}$Sr$_{2}$CaCu$_{2}$O$_{8+x}$, by contrast, the data might be
described by the pair breaking alone.
\end{abstract}


\end{frontmatter}

\section{Introduction}

In high-$T_{c}$ cuprate superconductors (HTCS) the microwave response
has been extensively investigated in order to get information, among
the others, on the symmetry of the order parameter, on the vortex
parameters such as the vortex viscosity and pinning frequency
\cite{golos,tsuchiya} and on the temperature dependence of the
superfluid fraction (via the measurement of the temperature dependence
of the London penetration depth \cite{hardy}).  In the vortex state
the response has been analysed mainly in terms of vortex motion, often
assuming that the field dependence of the quasiparticle (QP) and the
superfluid (SF) could be neglected.  However, there are some
experimental results that point to a very relevant role of QP/SF in
the determination of the high-frequency response in the vortex state.
In particular, in Bi$_{2}$Sr$_{2}$CaCu$_{2}$O$_{8+x}$ (BSCCO) it has
been shown that, at low enough temperatures, the imaginary
sub-terahertz conductivity is entirely due to the strong field
dependent superfluid depletion in a superconductor with lines of nodes
in the gap \cite{mallozzi}.  Moreover, we have recently shown
\cite{silvaSmBCO} that in SmBa$_{2}$Cu$_{3}$O$_{7-\delta}$ (SmBCO)
thin films the 48 GHz microwave resistivity is made up by two markedly
different contributions, namely a linear (in the field induction $B$) 
term, due to the vortex motion, that appears in the real part of the
microwave resistivity only (due to the fact that at high frequencies
the vortex motion is purely dissipative), and a sublinear one,
$\sim\sqrt{B}$, which reflects the strong pair breaking outside the
vortex cores \cite{volovik}.  Noticeably, in SmBCO the field-dependent
QP/SF response is detected up to very close to $T_{c}$.  Aim of this
paper is to present a compared study of the microwave response of
SmBCO, YBa$_{2}$Cu$_{3}$O$_{7-\delta}$ (YBCO) and BSCCO in the vortex
state in moderate fields, for temperatures not too far from $T_{c}$.
By assuming that the linear term in the real resistivity is due to
vortex motion, we can directly extract the complex QP/SF conductivity.
We find that YBCO and SmBCO behave similarly, and that the description
based on flux flow + SF depletion is fully consistent with the data.
By contrast, this framework does not apply to BSCCO, where it seems
that vortex motion does not play any relevant role in our temperature
and field ranges.  In this compound, the conductivity seems to be due
solely to field induced QP/SF variations.\\

\section{Experimental}

The samples under investigation are highly oriented, 200 nm thin films
of YBCO, SmBCO and BSCCO. Preparation details and crystallographic
characterization are reported elsewhere \cite{cucolo,camerlingo}.
Critical temperatures are 87, 86.5 and 89 K respectively, as estimated
from the crossing of the microwave real and imaginary fluctuation
conductivity \cite{fluct} (the typical $\pm$0.5K uncertainty of this
method is inessential for the purposes of the present paper).  The
$(a,b)$-plane complex resistivity $\tilde{\rho}$ is measured with a
metal cylindrical resonant cavity, operating in the TE$_{011}$ mode at
$\nu_{0}$=48.2 GHz in the end-wall-replacement configuration.  A
moderate magnetic field, aligned with the c axis, is swept at each
temperature from 0 to $B\simeq \mu_{0}H=$0.8 T (we expect that the
approximate equality between applied and internal field can be
inaccurate at very low fields only).  In the temperature range here
explored, $T>$70 K, the thin-film approximation is justified
\cite{film} and measurements of the field-induced shifts of $Q$ factor
and $\nu_{0}$ yield the field-induced variations of the complex
resistivity, $\Delta\tilde{\rho}(B)$, at fixed temperatures.  The
determination of $\Delta\tilde{\rho}(B)$ requires only the knowledge
of geometrical factors; absolute values of $\tilde\rho$ require the
cavity calibration.  In the following we will use normalized
resistivities, defined as $\tilde r=r_1+\mathrm{i}
r_2=\frac{\tilde\rho}{\rho_{0}}$, with $\rho_{0}=\rho_{1}(100K)$, and
the corresponding normalized field variations $\Delta\tilde r=\tilde
r(B)-\tilde r(0)$.  Experimental results are as follow: the two REBCO
samples show a similar behaviour in the entire temperature range
examined up to near $T_{c}$.  Measurements, plotted as a function of
$\sqrt{B}$, are reported in fig.1 for a selected set of field sweeps
at different temperatures for the YBCO sample.

\begin{figure}[htb]
\includegraphics [width=6.5cm]{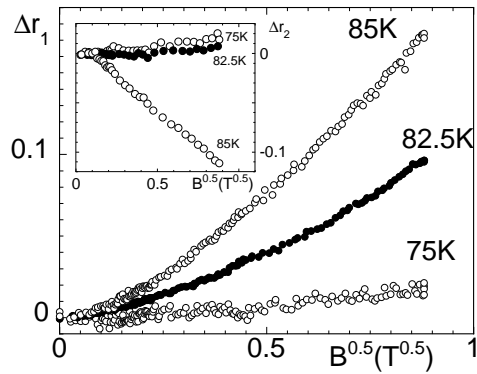}
\caption{Normalized complex
resistivity vs $\sqrt B$ in YBCO at selected temperatures. Main
panel: $\Delta r_{1}$; inset: $\Delta r_{2}$.}
\label{fig1}
\end{figure}

It can be noted that $\Delta r_{2}$ is well approximated by a straight
line, which corresponds to a $\sim \sqrt{B}$ dependence.  Moreover,
$\Delta r_{2}$ changes from positive to negative as the temperature
increases, but without changing the functional dependence on the
magnetic field.  On the other hand, $\Delta r_{1}$ is always positive:
an additional upward curvature in the data plotted as a function of
$\sqrt{B}$ indicates the presence of both a square root and a linear
term in the $B$ dependence.
\begin{figure}[htb]
\includegraphics [width=6.5cm]{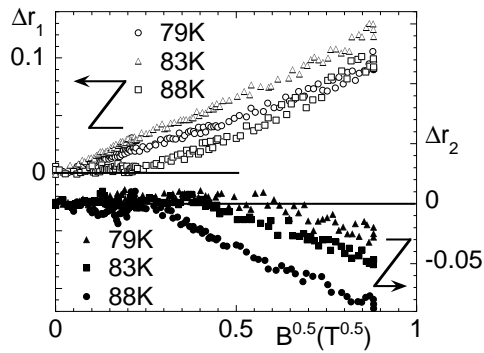}
\caption{Normalized complex resistivity vs $\sqrt B$ in BSCCO
at selected temperatures.  Open symbols: $\Delta r_{1}$; full symbols:
$\Delta r_{2}$.} \label{fig2}
\end{figure}

We now examine the measurements in BSCCO, where significant
differences with respect to the behaviour of REBCO are observed (fig.
2). In the temperature region above 78 K $\Delta r_{2}(B)$ is always
decreasing, but shows a $B$ dependence made up of a combination of a
linear term and a square-root one.  $\Delta r_{1}(B)$, on the other
hand, has essentially a $\sqrt{B}$ dependence up to 85K, where it
starts to change to a linear $B$ dependence, which holds up to 95 K
(well above $T_{c}$).  It appears that the behaviour in BSCCO is
somewhat reversed with respect to SmBCO and YBCO, exhibiting a
dominant $\sqrt{B}$ term in the real part of the resistivity and a
significant $\sim B$ term in the imaginary part.  We will discuss in
the following section the indications that can be gained from this
experimental behaviour.

\section{Discussion}

As extensively discussed in \cite{silvaSmBCO}, at high enough
frequencies, low fields, and not too close to $T_{c}$, the
vortex motion and QP/SF contributions to the microwave complex
resistivity are additive, and one can write $\tilde 
r(B,T)=\tilde r(0,T)+\Delta\tilde r(B,T)$, with:

\begin{equation}
\label{empirical1}
\Delta r_{1}=\Delta r_{1}^{vm}+\Delta
r_{1}^{qpsf}\simeq b_{1}(T)B+a_{1}(T)\sqrt{B}
\end{equation}

\begin{equation}
\label{empirical2} \Delta r_{2}=\Delta r_{2}^{vm}+\Delta
r_{2}^{qpsf}\simeq b_{2}(T)B+a_{2}(T)\sqrt{B}
\end{equation}
where we have explicitly indicated the vortex motion and the QP/SF 
terms, the latter approximate equality holds as an expansion for small 
fields, and we have cosidered the fact that the superfluid fraction 
decreases as $\sqrt{B}$ in a superconductor with lines of nodes 
\cite{volovik}.
In this framework, $b_{1,2}$ are related to vortex 
motion. 
We note that the knowledge of $\tilde r(0,T)$ and of the vortex motion terms 
$b_{1,2}$ allows for the inversion of the QP/SF terms and the 
determination of the field-dependent conductivity due to to QP and SF 
only.\\
We will discuss the data in the light of
Eq.s \ref{empirical1},\ref{empirical2}.  Due to their similarity, we
discuss first the results in YBCO and SmBCO, and we successively
consider the results in BSCCO.
The experimental results obtained for the two REBCO samples find a
very satisfactory explanation in terms of
Eq.s \ref{empirical1},\ref{empirical2} in the free-flux-flow limit (see
\cite{silvaSmBCO} for an extended data presentation and discussion in
SmBCO).  Therefore, once the coefficient $b_{1,2}(T)$ are known, the
QP/SF conductivity can be extracted by subtracting the linear terms
from the total resistivity and inverting the relation.  The result is
given in fig.  3 for the superfluid (imaginary) conductivity
$\sigma_{2}$ in SmBCO and YBCO: it can be clearly seen that
$\sigma_{2}$ shows a remarkable $\sqrt B$ trend, in agreement
with the prediction of the $\sqrt{B}$ dependence for the superfluid
density in a nodal superconductor.

\begin{figure}[htb]
\includegraphics [width=6.5cm]{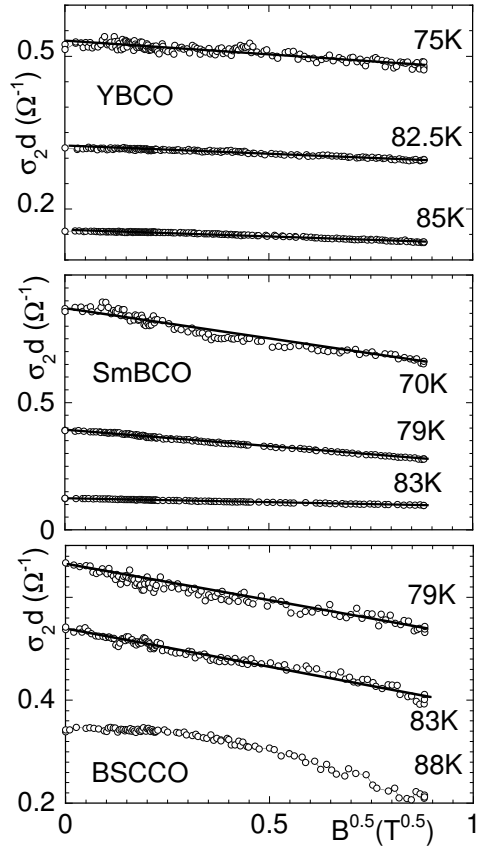}
\caption{Superfluid conductance ($d$ is the film thickness) vs $\sqrt
B$ at selected temperatures in YBCO, SmBCO and BSCCO (from top to
bottom).  The $\sim \sqrt{B}$ behaviour, as expected for a superconductor
with lines of nodes in the gap is ubiquitous (straight lines through 
the data points are guides for the eye). Only in BSCCO at high
temperatures the field dependence switches to linear, $\sim B$.}
\label{fig3}
\end{figure}
For the BSCCO sample we initially note that above $\sim$88 K $\Delta
r_{1}$ has evolved from a square root to a pure linear $B$ dependence.
Since the latter field dependence continues through and above the
critical temperature, we argue that at high temperatures the response
is dominated by fluctuations, which we will analyze in a future work.
We therefore consider the temperature range below 88K. We first note
that the imaginary resistivity is a decreasing function of the magnetc
field, thus ruling out possible strong pinning effects (we remind that
conventional vortex motion requires $\Delta r_{2}>0$ in the strong
pinning regime \cite{golos,cc}).  It would be then natural to apply
the framewok based on flux flow + pair breaking, as successfully
applied in REBCO. However, this approach does not appear to be
consistent: in fact, there is no evident linear term in the real part
of the field-induced resistivity variation, while a linear term is
present in the imaginary part.  This is somewhat puzzling in the frame
of a conventional vortex dynamics.  While complex vortex dynamics
cannot be definitely ruled out, in order to explore other alternative
scenarios we consider a somewhat opposite (and oversimplified) view:
we assume that the measured complex resistivity is entirely determined
by the field dependence of charge carriers conductivity, with no
relevant contribution from the vortex motion.  In this frame, the
measured $\tilde\rho(B)$ can be directly inverted to obtain
$\sigma_{2}$.  The result is given in fig.  3, lower panel, in which
$\sigma_{2}$ shows a remarkable $\sqrt B$ trend up to near $T_{c}$,
where it deviates upon entering the different (fluctuational) regime.
The total irrelevance of vortex motion contribution is clearly an
oversimplification, and more accurate models should take the vortex
motion into account.  Nevertheless, neglecting the vortex contribution
brings the QP/SF conductivity of BSCCO in close similarity to the one
obtained in YBCO and SmBCO.

\section{Conclusions}

We have reported measurements of the complex resistivity $\tilde\rho$
in a static applied field for three HTS materials, YBCO, SmBCO and
BSCCO. Both REBCO samples showed a $B$ dependence satisfactorily
explained as standard vortex dynamics added to a $\sqrt B$
pair-breaking term for the superfluid density, as expected for
superconductors with nodes in gap.  On the other hand, the BSCCO
sample showed a field-induced variation of $\tilde\rho$ inconsistent
with standard vortex dynamics.  By contrast, exploiting an oversimplified
model where no relevant vortex contribution is present, and the whole
response is due to the quasiparticle increase due to the strong
field-induced pair breaking in a nodal supeconductor, the obtained
superfluid conductivity $\sigma_{2}$ is found to be consistent with
theoretical expectations and with the $\sigma_{2}$ determined in
REBCO. In all cases it appears that field induced QP
excitations play an essential role, as witnessed by the clear
$\sqrt{B}$ dependence of $\sigma_{2}$.

\ack

This work has been partially supported by Italian MIUR under FIRB 
project ``Strutture semiconduttore/superconduttore per l'elettronica 
integrata''.

\end{document}